\documentclass[12pt]{article}
\title{
Test of the QCD vacuum with the sources in higher
representations }
\author{Yu.A.Simonov\\ State Research Center\\Institute of
Theoretical and Experimental Physics, \\ Moscow, Russia} \date{}
\newcommand{\be}{\begin{equation}} \newcommand{\ee}{\end{equation}}

\def\fun#1#2{\lower3.6pt\vbox{\baselineskip0pt\lineskip.9pt
\ialign{$\mathsurround=0pt#1\hfil
##\hfil$\crcr#2\crcr\sim\crcr}}}
 \newcommand{\lan}{\langle}
\newcommand{\ran}{\rangle}
 \newcommand{\llan}{\langle\langle}
\newcommand{\rran}{\rangle\rangle}

\begin{document}
\maketitle

\begin{abstract}
Recent accurate measurement \cite{1} \cite{2} of static potentials
between sources in various $SU(3)$ representations provides a crucial
test of the QCD vacuum and of different theoretical approaches to the
confinement. In particular, the Casimir scaling of static potentials
found for all measured  distances implies a strong suppression of
higher cumulants and  a high accuracy of the Gaussian stochastic
vacuum.  Most popular models are in conflict with these measurements.
 \end{abstract}


1.  An accurate measurement of static potentials between sources in
the eight  different representations of $SU(3)$ group made recently
in \cite{1} reveals a new quantitative picture of the QCD vacuum and
provides a crucial test of existing theoretical models. Other
measurements  of static interaction \cite{2} are in general agreement
with \cite{1}.

The most useful way to represent  static potentials $V_D(r)$ in
representations $D=3,8,6,15a, 10,27, 24, 15s$     is through the
complete set of field correlators in the framework of the Field
Correlator Method (FCM) \cite{3}:
\be
V_D(r)=-\lim_{T\to\infty}\frac{1}{T} \ln \lan W(C)\ran,
\label{1}
\ee
where Wilson loop $W(C)$ for the rectangular contour $C=r\times T$ in
the (\ref{3}\ref{4}) plane  has the cumulant expansion,
\be
\lan W(C)\ran = Tr_D exp \int_S \sum_{n=2,4...} (ig)^n\llan F(1)
F(2).. F(n) \rran d\sigma (1)... d\sigma (n)
\label{2}
\ee
Here $F(k)d\sigma(k)= F_{34}(u^{(k)}, x_0) d\sigma_{34}(^{(k)})$ and
the component $F_{34}(u,x_0)\equiv E_3(u,x_0)=
\phi(x_0,u)E_3(u)\phi(u,x_0)$, where $\phi$ is a parallel transporter
and $x_0$ is an arbitrary point on the surface $S$ inside contour
$C$; $Tr_D\hat 1=1$.

Dependence on $D$ enters in (\ref{2}) through the generators $T^a$,
since $F(k)= F^a(k) T^a(a=1,... N_c^2-1),$ and the main
characteristics of $D$ is the quadratic Casimir operator $C_D$:
$T^aT^a=\hat1 C_D$, so that the invariant square of the color charge
in the representation $D$ is $g^2C_D$.

One can now express the connected correlators (cumulants) in
(\ref{2}) via $C_D$ and $D$--independent averages as follows
(for more details  see last reference in \cite{3} and\cite{4}),
\be
Tr_D\lan F(1) F(2)\ran =C_D\frac{\lan F^a(1) F^a(2)\ran}{N_c^2-1},
\label{3}
\ee
$$
Tr_D\llan F(1) F(2) F(3) F(4)\rran= \frac{C^2_D}{(N_c^2-1)^2}
\{\lan F^a(1) F^a(2) F^b(3)F^b(4)\rran
$$
$$+
\lan F^a(1) F^b(2) F^b(3)F^a(4)\ran
-\lan F^a(1) F^a(2)\ran \lan F^b(3)F^a(4)\ran
$$
\be
+(1-\frac{N_c}{2C_D})
\lan F^a(1) F^b(2) F^a(3)F^b(4)\ran\}+ O(\frac{1}{N_c^2}).
\label{4}
\ee
Note that the arguments of $F(k)$ in (\ref{4}) and in (\ref{2}) are
ordered (e.g. clockwise, $u^{(1)}
< u^{(2)}
< u^{(3)}
< u^{(4)})$
and therefore the only  vacuum insertion is possible in the first term
on the r.h.s. of (\ref{4}) leading to the cancellation with the third
term: hence the correlator (\ref{4}) is a connected one vanishing at
large distances, $|
 u^{(1)}+
 u^{(2)}-
 u^{(3)}-
 u^{(4)}|\to\infty$.

 One can show  in a similar way that the $n$-th cumulant in
 (\ref{2}) contributes proportionally to $C^n_D$. As a result the
 static potential $V_D(r)$ has the expansion
 \be
 V_D(r)= d_DV^{(2)}(r) + d^2_D V^{(4)}(r)+...,
 \label{5}
 \ee
  where in notations of ref. \cite{1}
  $d_D=C_D/C_F$ and $C_F$
  is the fundamental Casimir operator, $C_F=\frac{N_c^2-1}{2N_c}$.
        The fundamental static potential contains perturbative
        Coulomb part $V_{Coul}$,  confining linear and constant
  terms.

  The  Coulomb part ,which is also obtainable from the perturbative
  component of the FC in (\ref{3}) 
  is now known up to two loops \cite{5} and is
  proportional to $C_D$. Therefore one may expect quartic
  contributions proportional to $C^2_D\sim d^2_D$, to the constant
  and linear terms, writing (\ref{5}) as
  \be
  V_D(r)= d_D V^{(2)}(r)+ d^2_D (\bar v_0^{(4)} + \bar \sigma _4 r).
  \label{6}
  \ee
  Here
   $\bar v_0, \bar \sigma_4$
   measure the contribution of the
  quartic cumulants to the constant term and string
  tension respectively.

  Now the  measurements of ref. \cite{1} allow to find
   $\bar v_0, \bar \sigma_4$
   from all 8 sets of data. To this end one forms 7 combinations
   $\zeta_D\equiv V_D(r) -d_D V_F(r)= d_D(d_D-1)  (\bar
   v_0^{(4)}+\bar \sigma  r)$.
   As a typical example one can take
   fundamental and adjoint potentials,at distances between 0.05fm
   and 1.1fm
    from  the data \cite{1} the $\chi^2$ fit  yields for $\bar
   v_0, \bar \sigma_4$
   \be
   \bar v_0^{(4)}=(- 0.6\pm 0.67)\cdot 10^{-3}  GeV
   \label{7}
   \ee
   \be
   \bar \sigma^{(4)}=(- 1.136\pm 0.69)\cdot 10^{-3} GeV^2
   \label{8}
   \ee

   The quality of the fit is reasonable, $\chi^2/N= 0.45, N=43$
    One obtains similar figures also for D=6,15a,10(while 3 higher
    representations do not yield additional information)suggesting
    that $\bar \sigma^{(4)}$is negative while $\bar v_0^{(4)}$ is compatible
  
     with zero,confirming in this way the parametrization (\ref{6}).
   This analysis
     demonstrates the phenomenon of the Casimir scaling, i.e.
   proportionality of static potential $V_D(r)$ to the Casimir operator
   $C_D$ with accuracy  better than  one percent.

   Physical consequences of the Casimir scaling are numerous and
   important.

   First of all, the sign and magnitude of quartic correction
   (\ref{7}),(\ref{8}) can be  understood in the FCM. Indeed the
   quartic term enters the potential $V_D$ with the factor $(-g^4)$,
   as compared to $+g^2$ for the quadratic (Gaussian) term. Secondly,
   one can estimate $\lan E^2_3\ran$ term from the standard gluonic
   condensate as follows:
   \be
   g^2\lan E^a_3 E^a_3\ran\sim \frac{4\pi^2}{12} (0.04\pm 0.02)
   GeV^4\sim (0.10\pm 0.06) GeV^4
   \label{9}
   \ee
   and take into account that
    the cumulant expansion in (\ref{2}) is
   actually  in powers of the parameter
   \be
   \xi\equiv g^2
\lan E^a_3 E^a_3\ran T^4_g
\label{10}
\ee
Here $T_g$ is the correlation length of the QCD vacuum;
for bilocal correlator it was
measured on the lattice \cite{6}, $T_g^{(2)}\sim 1 GeV^{-1}$.
With the use of (\ref{10}) one could expect that $\bar \sigma_4$
would be from 4 to 10\% of the standard string tension, $\sigma=0.2
GeV^2$ provided $T_g=T_g^{(2)}$.
The  value of $\bar \sigma^{(4)}$
 calculated in (\ref{8}) is at least 6 times
smaller and suggests that quartic correlation  length $T_g^{(4)}$ may
be  smaller than the Gaussian one, $T_g^{(2)} \sim 0.2 fm$.

This result means that the Gaussian Stochastic Model (GSM), suggested
in \cite{3} and successfully  used  heretofore in many applications
\cite{7}, can be more accurate than it was even expected, at least in
processes where string tension plays the most important role. On the
other hand, the smallness of quartic and higher contributions implies
a very specific picture of vacuum correlations.

Indeed the smallness of $T_g^{(4)} $ implies that color fields tend
to form compact white bilocal combinations $F^a(1) F^a(2)$ which are
almost noninteracting between themselves and therefore not
contributing to the higher connected correlators.  This
looks like
the  picture of small white dipoles made of fields $FF$ (or of vector
potentials $A_\mu A_\mu$ connected to $FF$ in the Fock--Schwinger or
contour gauge).

 One can also understand qualitatively the difference between $T_g^{(2)}$
and
$T_g^{(4)}$, since $T_g^{(2)}$ measures correlation length between
adjoint fields $E^a_3(x)$ and $E^a_3(y)$
in  the Gaussian correlator
$\lan E^a_3(x)E^a_3(y)\ran$, while $T_g^{(4)}$ refers to
the correlation of two white complexes, and should be connected to
the lowest glueball mass, $M_G\approx 2GeV$; hence $T_g^{(4)}\sim
1/M_G\sim 0.1 fm <T_g^{(2)}.$

Finally, the Casimir scaling imposes severe restrictions on existing
models of the QCD vacuum. For example, the center--symmetry
flux model was tested and ruled out in \cite{1}, since in the
original formulation it predicts  vanishing adjoint string
tension, while in the later modification -- the  fat vertex model
\cite{9} it is still far from the accurate  data \cite{1}.

Next one should mention models of the abelian projected vacuum which
fail to provide Casimir scaling \cite{11}, at least in the simplest
version \cite{12}.

Consider now the dilute instanton gas model and
the $SU(2)$ group. The instantons
 may be present in the confining
vacuum as an important source of chiral symmetry breaking.  Then
the Casimir scaling \cite{1} imposes a strict bound on the
admixture of instantons in the QCD vacuum. Indeed, insertion in
(\ref{2}) and (\ref{4}) of the instanton field strength $gF^a_{34}
(x, z) =\frac{4\delta_{a3}\rho^2}{[(x-z)^2+\rho^2]^2}$, $(\rho$ --
instanton size, $z$ -- its position,   contribution of
parallel transporters
is neglected for simplicity since it gives for bilocal
correlators a reduction of 20-30\%, see \cite{Ilg} for details)
yields  the
following  expression for the quartic contribution to the static
potential ($r< \rho$) (in $SU(2)$
case)
  \be
V_D^{(4)} (r\to 0) = -\frac{N}{V}\frac{r^4}{\rho}\>
\frac{3C_D^2 - C_D}{15}\frac{\pi^6}{320}
 \label{11}
\ee
 where $\frac{N}{V}$ is the density of instanton in the vacuum.
Inserting here the values of $v_0^{(4)}$ above(Eq(\ref{7})) one gets the
following
bounds 
on the density of instantons
\be
 \frac{N}{V}\leq 0.17 \; fm^{-4},
 \label{12}
  \ee
which is much smaller than normal instanton density of $1 fm^{-4}$.

With such density  the role of instantons in chiral symmetry breaking
and other effects would be negligible.

A more stringent bound can be obtained from the quartic string tension 
generated by instantons . However 
the nonzero value of $\bar \sigma_4$ for instantons
does not imply
confinement. One should take into account that at large
distances the sum of all partial string tensions
$\sum_{n=2,4,...}\bar \sigma^{(n)}$ for the dilute instanton gas
vanishes \cite{Sim},\cite{13}.

In previous discussion we have ignored the fact that at large
enough distances the adjoint charges are screened by the vacuum
gluons,  and the limiting value of the  adjoint potential is equal to
the doubled  gluelump mass, $2M_{gl}$. This leads to an estimate of
the screening distance $r_0$ from the relation;  $V_{adj}(r_0)=2
M_{gl}$ where $M_{gl} $ from \cite{gr} is around 1.4 GeV  and
therefore $r_0\approx 1.4 fm$, which is beyond the distance
where Casimir scaling was measured in \cite{1}.

Thus the Casimir scaling is a  stringent test for all models
considered and displays a strong suppression
of quartic and higher connected correlators,hence
supporting a good accuracy of the GSM.
 At this point one may wonder how the negligible small higher 
 correlators are compatible with the screening of the adjoint
 potential at $r\leq r_0$,which is not seen in the cumulant expansion
 (\ref{2}).The solution to this problem was suggested in \cite{gr},
 where the screening terms have been identified as an addition to the
 Wilson loop,with the small coefficient proportional to $N_c^{-2}$.
 The corresponding term actually comes from the two- and more
 Wilson loop averages,and therefore has a perimeter rather 
 than area-law behaviour.Hence sreening cannot be seen in the
 one-Wilson-loop expansion (\ref{2}), and the transition due to the
 properties of the definition (\ref{1}) of the static potential 
 occurs rather sharply in large  $T$  limit.

This work was partially supported by the joint grant RFFI-DFG,
96-02-00088G. The author is grateful to Dr. G.Bali for sending his
lattice data and for correspondence, to A.M.Badalian, D.I.Diakonov
M.I.Polykarpov and G. 't Hooft for useful discussions, 
 and V.I.Shevchenko for
discussions and help in calculations.

\end{document}